\documentclass[12pt]{article}
	
    \makeatletter
    \renewcommand\section{\@startsection {section}{1}{\z@}%
                                       {-3.5ex \@plus -1ex \@minus -.2ex}%
                                       {2.3ex \@plus.2ex}%
                                       {\normalfont\fontfamily{ptm}\fontsize{16}{19}\bfseries}}
    \renewcommand\subsection{\@startsection{subsection}{2}{\z@}%
                                         {-3.25ex\@plus -1ex \@minus -.2ex}%
                                         {1.5ex \@plus .2ex}%
                                         {\normalfont\fontfamily{ptm}\fontsize{14}{17}\bfseries}}
    \renewcommand\subsubsection{\@startsection{subsubsection}{3}{\z@}%
                                        {-3.25ex\@plus -1ex \@minus -.2ex}%
                                         {1.5ex \@plus .2ex}%
                                         {\normalfont\normalsize\fontfamily{ptm}\fontsize{14}{17}\selectfont}}
    \makeatother
	
	\usepackage{amsmath}
	\usepackage{graphicx}
	\usepackage{enumerate}
	\usepackage{xcolor}
	\usepackage{natbib}
	\usepackage{url}
    \usepackage{hyperref}
    \usepackage{url}
    \usepackage{booktabs}
    \usepackage{nicefrac}
    \usepackage{microtype}
    \usepackage{mathtools}
    \usepackage{algorithm}
    \usepackage{multirow}
    \usepackage{lscape}
    \usepackage{amsfonts}
	
	\date{}
	
\begin{document}
		
		\def\spacingset#1{\renewcommand{\baselinestretch}%
			{#1}\small\normalsize} \spacingset{1}
		
		\title{\textbf{Towards Robust Representations of Limit Orders Books for Deep Learning Models}}
		
		\author{
		Yufei Wu\thanks{Yufei Wu is a senior AI Research Associate at J.P. Morgan AI Research in London, United Kingdom. (yufei.wu@jpmorgan.com)}, 
		Mahmoud Mahfouz\thanks{Mahmoud Mahfouz is an AI Research Lead at J.P. Morgan AI Research and a PhD student at Imperial College London, United Kingdom. (mahmoud.a.mahfouz@jpmchase.com)}, 
		Daniele Magazzeni\thanks{Daniele Magazzeni is an AI Research Director at J.P. Morgan AI Research in London, United Kingdom. (daniele.magazzeni@jpmchase.com)}, 
		Manuela Veloso\thanks{Manuela Veloso is Head of J.P. Morgan AI Research in New York, United States of America. (manuela.veloso@jpmchase.com)}
		}
		
		\maketitle
		
		\begin{abstract}
    
    The success of deep learning-based limit order book forecasting models is highly dependent on the quality and the robustness of the input data representation. A significant body of the quantitative finance literature focuses on utilising different deep learning architectures without taking into consideration the key assumptions these models make with respect to the input data representation. In this paper, we highlight the issues associated with the commonly-used representations of limit order book data from both a theoretical and practical perspectives. We also show the fragility of the representations under adversarial perturbations and propose two simple modifications to the existing representations that match the theoretical assumptions of deep learning models. Finally, we show experimentally how our proposed representations lead to state-of-the-art performance in both accuracy and robustness utilising very simple neural network architectures.
    
\end{abstract}
    	
    	\newpage
        \tableofcontents
        \newpage
	    
	    \section{Introduction} \label{introduction}

    Limit order books (LOBs) are used by financial exchanges to match buyers and sellers of a particular instrument and act as an indicator of the supply and demand at a given point in time. It can be described as a self-evolving process with complex spatial and temporal structures revealing the price dynamics at the microstructural level. Market making, optimal execution and statistical arbitrage strategies, all require a good understanding of the LOB and its dynamics. \\
    
    \begin{figure}[!htb]
        \centering
        \includegraphics[width=\textwidth]{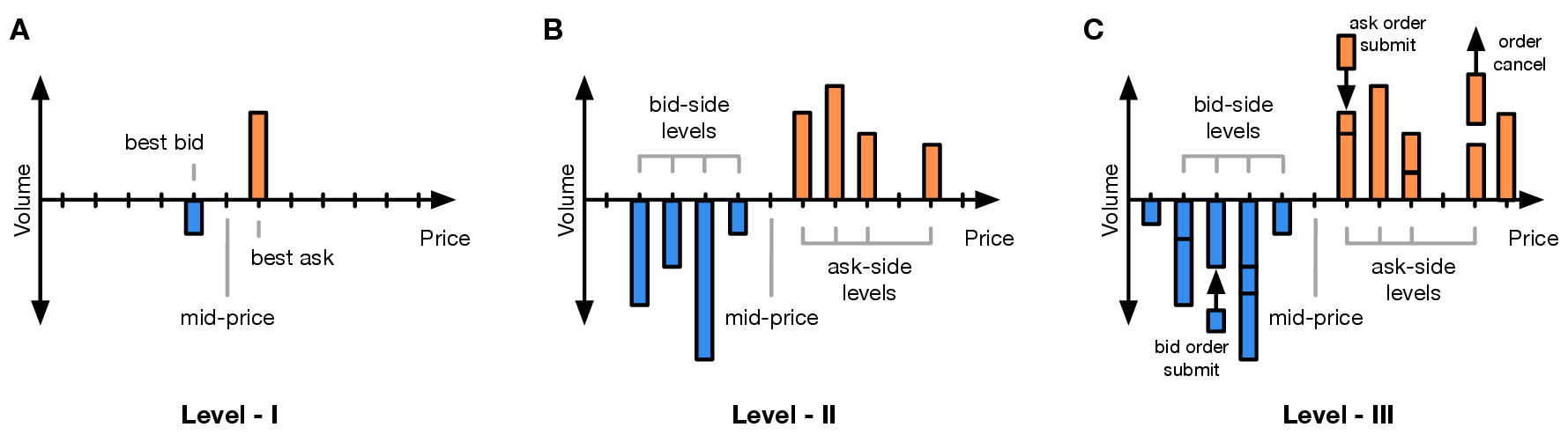}
        \caption{Limit order book data in different degrees of granularity.}
        \label{fig:lob}
    \end{figure}
    
    A \emph{limit order} is a buy (bid) or sell (ask) order for a certain quantity of an exchange-traded instrument at a certain discrete price. The minimum price increment change is called the tick size which is \$0.01 for US stocks on NASDAQ for example. A limit order remains in the limit order book until it is executed or canceled. A \emph{market order}, on the other hand, is an order for a specific quantity to be matched at the best available price. This is the lowest price of ask limit orders (best ask prices) or the highest price of bid limit orders (best bid prices) pending in the market. A market order typically ensures immediate execution with a compromise on the price. In these order-driven markets, all explicit order actions are captured by exchange messages, which are then used to generate the limit order book. Over time, the limit order book gets updated continuously with order placements, cancellations, and executions. Limit order book data come in different degrees of granularity (see figure \ref{fig:lob}), including \emph{Level-I} data providing the best bid/ask prices and volumes, \emph{Level-II} data providing deeper insights with price and aggregated volume information across a certain number of price levels and \emph{Level-III} data containing the non-aggregated orders placed by market participants. \\

    The use of algorithmic trading strategies and the digitisation of exchange activities has made available a tremendous amount of LOB data for practitioners and researchers allowing them to study the market dynamics using various data-driven approaches. This led to a surge in interest for big data applications in the financial markets and machine learning (especially deep learning) models have become a trend in the quantitative finance domain \citep{buehler2019deep,wiese2020quant}. \\
    
    One of the classical and popular tasks using limit order book data is the short-term price forecasting task, which is studied extensively in the academic literature and is valuable from the commercial perspective for markets including equities, currencies and commodities. Many of the existing price forecasting models in the literature use Level-1 data and rely on the time-related features to predict the future price movements in an auto-regressive way. Recent studies \citep{tsantekidis2017using,tsantekidis2017forecasting,tran2018temporal,zhang2019deeplob,mahfouz2019importance,sirignano2019deep,tsantekidis2020using} focused on using machine learning (including deep learning) models on Level-2 data. Due to the richer features in the Level-2 data than the Level-1 data and more powerful computational models, deep learning forecasting models showed state-of-art performance beyond the capacity of conventional models. \\
    
    Neural networks with convolutional layers are commonly used due to their success in areas such as computer vision. However, most limit order book-based deep learning models use the convolutional layers in a fundamentally incorrect way, leading to seemingly-working but vulnerable and unreliable models. To explain this, we need to start from the basics of information and machine learning. Original real-world signals come in various forms and there are multiple ways to digitise the raw information for computers to process. For deep learning models, input data need to be organised as tensors. A tensor is a multi-dimensional array with a uniform type for each entry. A 1-D tensor is a vector and a 2-D tensor is a matrix. All future learning processes rely on the organisation scheme of tensors. A deep neural network can be viewed as a combination of a feature extractor and a predictor. The feature extractor converts the input data into a high-level representation by extracting useful and meaningful features. If the feature extraction techniques are applied manually to the raw data, this conversion from raw data to feature vectors is referred to as \emph{feature engineering}. This requires a good and comprehensive domain knowledge to make sure the extracted features match the learning task. By contrast, the feature extraction in deep learning models, referred to as \emph{feature learning} is an automated approach to discover an optimal representation for the data. The major difference between feature engineering and feature learning is whether the representation is formed in a purely data-driven fashion and the former is more popular in conventional non deep learning approaches. On the other hand, the performance of the predictor is heavily influenced by how the features are designed or learned \citep{bengio2013representation}. For neural networks, the feature learning and the predictor are combined within the network structure and are trained together towards the same target function. In this case, the original representation of the limit order book data, \textit{i.e.} the input representation to the neural networks, becomes the foundation of the entire model and needs to be compatible with the feature learning structure.

    \subsection{Related Work}
    
        Early work using high frequency limit order book data used handcrafted features combined with learning models such as support vector machines (SVMs) to indicate future price movement \citep{kercheval2015modelling}. As deep learning methods became more popular, researchers explored deep neural networks to solve large-scale problems in the context of price prediction using multi-layer perceptrons \citep{mahfouz2019importance}, recurrent neural networks \citep{sirignano2019universal}, long short-term memory \citep{tsantekidis2017using}, convolutional neural networks \citep{zhang2019deeplob,zhang2018bdlob} and recently sequence to sequence (Seq2Seq) networks\citep{zhang2021multi}. See \cite{lim2021time} for more detailed review of time series forecasting with deep learning models. \\ 
    
        Although deep learning is becoming more widely used in finance, the literature rarely discusses the representation schemes of the financial datasets. A vast majority of the machine learning models, including those mentioned above, and benchmark datasets arrange the LOB data in a price level-based representation (e.g. \citep{ntakaris2018benchmark,huang2011lobster}). However, a LOB representation which is efficient and convenient from the perspective of human understanding does not necessarily mean it is an appropriate representation scheme for machine learning models to learn features from. The study of the importance of robust data representation, the criteria for evaluating the quality of the representation, and the variety of methods for learning these representations is studied extensively in the machine learning literature with \citep{bengio2013representation} providing a survey of these methods. \\ 
    
        In our work, we focus on the representation of financial market microstructure data. \citep{bouchaud2018trades,abergel2016limit} study the structure and empirical properties of limit order books and provide a set of statistical properties (referred to as \textit{stylized facts}) using NASDAQ exchange data. On the other hand, \citep{lehalle2018market} discusses the practical aspects and issues of market structure, design, price formation, discovery and the behaviour of different actors in limit order book markets. A significant amount of research in recent years focused on applying deep learning models on limit order book data for the purposes of price forecasting or price movements classification.
    
    \subsection{Our Contributions}
        
        Our work, to the best of our knowledge, is the first bringing adversarial robustness to LOB representations and pointing out the critical flaws of the commonly-used LOB representation with machine learning models. Based on these concerns, we present desiderata of LOB representations and propose new LOB representation schemes which lead to more effective and robust machine learning models. The experimental results confirm our concerns about the current level-based LOB representation as well as machine learning models designed based on this representation scheme while \textbf{new representations we propose outperform state-of-art methods in both accuracy and robustness}. This emphasises the importance of compatibility of different aspects of a machine learning model from data representation to model structure.

        \section{Current Limit Order Books Representations}
\label{sec:representation}

    For limit order books, there exists a number of possible price ticks allowing orders to be placed and those with no orders placed in them lead to empty ticks. This is especially prevalent in small-tick instruments where a relatively large proportion of price ticks can be left unoccupied. There are two common ways to represent such information with one being a plain vector and the other in a compressed form. The vector representation displays the information as vectors (or matrices if multi-dimensional) with zero entries representing empty ticks. Due to the common sparsity in those vectors, one can also compress the representation by describing each non-zero entry with a pair of indications. Usually, the first element in pairs stands for the index of one of the non-zero sparse vector entries while the second element is the corresponding value. \\
    
    From the data structure perspective, compressed or normal vector representations demonstrate different characteristics. One obvious difference is the storage size - the amount of storage space a compressed representation occupies is $2M$ where $M$ is the number of non-zero entries in the original vector, which would be more efficient as long as $M < N/2$ where $N$ is the length of sparse representation. Furthermore, each element in the vector representation stands for signals from the same source, which is no longer valid in the compressed representation. A compressed representation includes tuples of (source index, source value).  Elements inside a tuple are tied together but tuples are independent among each other and the compared representation is not unique\footnote{$\vec{\tilde x} = [(2,2), (6,1), (5,3)]$ is also a valid compressed representation, although usually the orders of tuples are arrange with some form of index ranking.}. Any disentanglement or distortion to this would result in invalid representations. Also, with compressed representations, any processing, e.g. projection, addition, multiplication, distancing etc., applied to indices will not lead to meaningful results. 
    
    \subsection{Vector Representation}
    
        Limit order books can be represented as vectors. In this format, each entry stands for a certain price level which are arranged monotonically and the value at each entry indicates the accumulated volume (can be zero) at this price level. For price forecasting tasks, limit orders near the mid-price play an important role compared with orders placed far away from the mid-price. Thus, we are more concerned about limit order information near the mid-price. To ensure consistency, previous work \citep{abergel2016limit,abergel2020algorithmic} provide an approach to represent order book snapshots by including $W$ price levels on the bid and ask sides. The representation requires the best ask price $p_a^1(t)$ and best bid price $p_b^1(t)$ and uses a pair or vectors:
        \begin{equation*}
            (\vec a(t), \vec b(t)) = ([a^1(t),...,a^W(t)],[b^1(t),...,b^W(t)]),
        \end{equation*}
        where $a^i(t)$ is the volume at $i$ ticks away from $p_b^1(t)$ and $-b^i(t)$ is the volume at $i$ ticks away from $p_a^1(t)$. This representation requires 2 additional dimensions ($p_a^1(t)$, $p_b^1(t)$) to anchor the vectors. Compared with the compressed representation of LOBs, this representation scheme avoids some drawbacks. First, all numbers in this representation are volumes instead of volume-price couples, which avoids the risks of invalidity if disentanglement happens in future black-box models. This representation is also spatially homogeneous since distances between adjacent elements (spatially) are equal to the tick size. Thus, it doesn't have the issue of level shifts when empty ticks are filled. Because it includes all the empty ticks within the scope of vision, this representation may look sparse, \textit{i.e.} with a considerable amount of zero elements in the representation.
    
    \subsection{Compressed Representation}
    
        The representation for limit order books which is commonly used as inputs to most machine learning models for LOB predictions and financial benchmark datasets contains each snapshot as a compressed representation. Each data sample is in the format of $ \vec x \in \mathbb{R}^{T\times 4L}$, where $T$ is the duration of historical LOB snapshots and $L$ indicates the number of (non-zero) price levels included in the representation on both ask and bid sides. The compressed expression can be written as
        \begin{equation}
            \vec x_{t} = \left\{p_{a}^{i}(t), v_{a}^{i}(t), p_{b}^{i}(t), v_{b}^{i}(t)\right\}_{i=1}^{L},
        \end{equation}
        where $p_{a}^{i}(t)$, $p_{b}^{i}(t)$ are the ask and bid prices for price level $i$ and $v_{a}^{i}(t)$, $v_{b}^{i}(t)$ are the ask and bid volumes respectively. Reflected in the data format, $4L$ is the length of the vector representing each snapshot of the LOB at a certain time point.\\
        
        The compressed representation has some particular characteristics from a data representation perspective. The most intuitive one is that the price and volume for each LOB level are tied together - any disentanglement or distortion to this would result in invalid representations. In addition, information on each price level is ranked in an ascending (ask side) or descending (bid side) order in the compressed expression. Due to this ranking mechanism, the spatial structure across different levels is not homogeneous since there is no assumption for adjacent price levels to have fixed intervals, only a monotonic order is guaranteed. This also implies that representations can be altered dramatically due to occasional shifts in price levels - the previous best bid/ask data can suddenly shift to second best bid/ask channel if a new order is placed with a better price. \\
        
        These characteristics may not be a big problem for human understanding but they highlight the drawbacks of this LOB representation when treated as inputs to machine learning models. Firstly, it is fundamentally assumed in machine learning that signals from the same channel (input dimension) are from the same source. In this case, `level' is a manually defined concept based on the price ranks of current orders in LOB, which can hardly be treated as the same source especially when the information of a level shifts to the channel of another level due to certain actions. Secondly, homogeneous relationship is a basic assumption for convolutional neural networks (CNN) due to the parameter sharing mechanism. Thus, the heterogeneous features of LOB data representation may reduce model robustness when learning with CNN models. Furthermore, the way how the information is organised from multiple levels makes it vulnerable to perturbations - a small perturbation would leads to shift of price levels and thus the representation would be affected dramatically due to this shift.
        \section{Risks of Representations under Adversarial Perturbations}
\label{sec:risk}

    Adversarial perturbations are common and inevitable in most real-world systems. These can happen in any component of the data processing chain. Perturbations can be very subtle but still significantly harmful to sophisticated machine learning systems. If such systems are built with high complexity but without considering risks under possible perturbations during the model design and development phase, it would lead to instability and vulnerability in the post-deployment phases when facing real-world conditions. In the financial domain, it is important to ensure that the potential risks are acknowledged and controlled before model deployment to ensure systematic reliability. One way to examine this is by designing hypothetical conditions which might not have occurred before but can possibly happen in the future. Especially, when we have a relatively good understanding of the drawbacks of a certain system based on its design, we can develop scenarios with \emph{adversarial perturbations} to better identify risks. Unlike adversarial samples, which are targeted specifically on individual samples, we refer adversarial perturbations to more general conditions that can be applied to all inputs in the same manner. \\ 
                
    In this paper, we propose an adversarial perturbation scheme, which is possibly caused by valid market operations -  limit order placement in the market. This perturbation scheme assumes that the limit order book is perturbed by orders of small sizes at empty price levels beyond the best ask/bid prices. This hypothesised situation ensures no change is made to the mid-price before and after the perturbation to make sure the prediction labels are not affected. In some LOB data for equities, the price difference between adjacent price levels is sometimes larger than the tick size (the minimum price increment change allowed). This is especially prevalent in small-tick stocks and can result in the entire LOB shifting even if a small order of the minimum allowable size is placed at a price in between the existing price levels. \\
        \begin{figure*}[!tb]
            \centering
            \includegraphics[width=\textwidth]{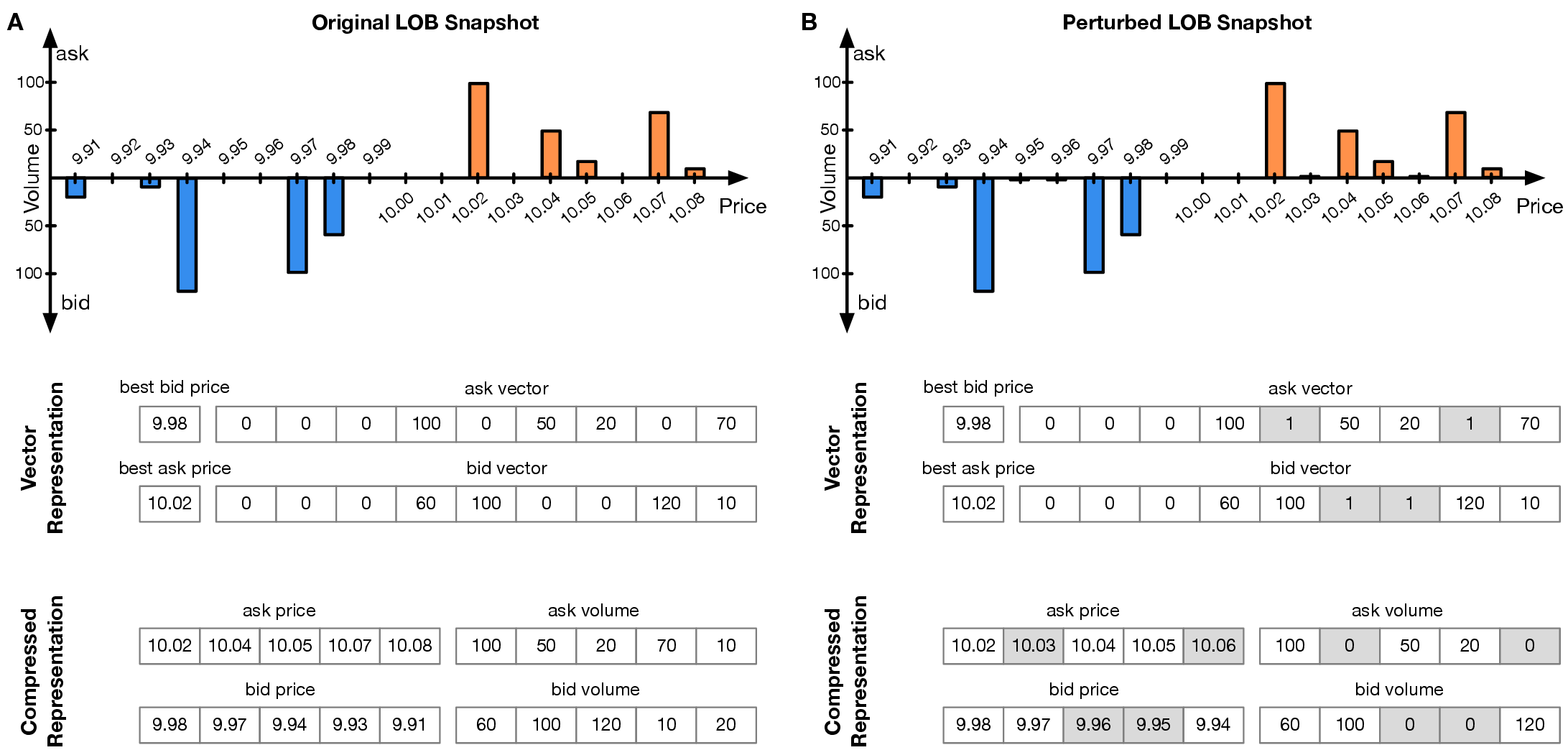}
            \caption{(\textbf{A}) Original LOB data with 5 levels on the ask and bid sides without perturbation. (\textbf{B}) LOB data with 5 levels after data perturbation. grey blocks represent perturbation orders with order volume = 1. }
            \label{fig:perturbation}
        \end{figure*}
    
    We illustrate the influence of adversarial perturbation on representations with a synthetic LOB example as shown in Figure \ref{fig:perturbation}. Figure \ref{fig:perturbation} (A) shows the synthetic LOB snapshot with price levels for both the ask and bid sides of the LOB before any perturbation. The tick size is assumed to be 0.01 and the minimum order size present in our data is 1. In this LOB snapshot, the mid-price is 10.00 with bid-ask spread equals to 0.04. We can observe some price levels where no orders are placed, such as 10.03, 10.06 in the ask side and 9.96, 9.95 in the bid side. Two LOB representation methods are demonstrated for the same LOB snapshot, which is vector representation and compressed representation as introduced in Sec. \ref{sec:representation}. Both representations are restricted to 20 dimensional. The vector representation encodes price and volume information in the window (9.93-10.07) near the mid-price while the compressed representation covers from 9.91 to 10.08.\\
    
    To perturb this LOB data, one can place orders with the allowed minimum order size to fill some empty price levels. These minimum size orders may seem to be not influential since 1) they do not affect the mid price, 2) their volumes are small. As shown in Fig. \ref{fig:perturbation} (B), 4 limit orders are placed with size equals to 1 to price level 9.95, 9.96, 10.03, 10.06 respectively but the LOB snapshot demonstrates subtle difference visually compared with Fig. \ref{fig:perturbation} (A). For vector representation, due to plain reflection of the snapshot, it adds 1's to the corresponding zero entries in its representation while other non-zero ones remain unchanged from perturbation. However, the compressed LOB representation, the most commonly used one in real-world tasks, changes dramatically after this perturbation. Approximately half of the original price level information is no longer visible after perturbation (e.g. ask-side L4, L5 information pre-perturbation is no longer included in the representation after perturbation) and while the rest are preserved, they are shifted to different levels in the LOB representation (e.g., the ask-side L2 appears in ask-side L3 after perturbation). \\ 
    
    From this example, we can deduce that the vector representation is more robust under certain perturbations compared with the compressed representation. Intuitively, this adversarial perturbation impacts the compressed representation from the machine learning point of view in multiple ways. Firstly, it shifts the 20-dimensional input space significantly. For example, the L1-norm distance between these two 20-dimensional vectors before and after perturbation is near 400 whereas actually the total volume of orders applied is only 4. By contrast, the L1-norm distance of vector representation before and after perturbation is 4. This means: 1) the compressed representation scheme based on price levels does not bring local smoothness and 2) it amplifies the perturbation from the market activity to the digital description of the market dramatically and unreasonably. Furthermore, this adversarial perturbation may potentially narrow the scope of vision of machine learning models to \textit{observe} the LOB. As shown in Fig. \ref{fig:perturbation}, the visible scope for compressed representation narrows down to 9.94-10.06 from the original 9.91-10.08.

        \section{Desiderata of LOB Representations}
\label{sec:desiderata}

    Considering market activities that take place in the time scale of milliseconds or even nanoseconds, both storing and processing limit order book data faces the obstacles of dealing with big data problems. The good news is, since the limit order book is an accumulation of market activities, one can store the message information i.e. changes of the limit order book, instead of the entire order book to reduce the storage cost. Thus, the main challenge brought by the big data raises when applying machine learning models on limit order book data. \\
    
    In this paper, we propose some desiderata guiding the preparation and representation of limit order book data for machine learning models. These desiderata come not just from the perspective of whether the representation is a meaningful, comprehensive and efficient way to reflect the original information. They are driven from the machine learning point of view, concerning whether this representation is compatible and appropriate for the machine learning model to be used in real tasks. Note that, data can be represented differently in storage, transition or analysis and our desiderata only applies on the representation directly fed to machine learning models as inputs. 
    
    \begin{itemize}
    
        \item \textbf{Clarity}: The data representation should include a clear and ideally simple definition of encoding and decoding transformation towards other representation schemes. It should also avoid non-trivial methods to detect the validity of data samples. 
        
        \item \textbf{Efficiency} The entire limit order book may contain hundreds of price levels with a large range of price levels. A complete representation including all price levels is not always necessary for all the tasks. Thus, an appropriate region of interests needs to be placed to the limit order book to reach a balance between complexity and performance. For example, limit order books snapshot can be represented with extremely sparse vector including all price ticks that appeared historically. This kind of representation is complete, easy to understand but very inefficient both in storage and in computation. A LOB representation should organise data in an efficient manner to reduce the \emph{curse of dimensionality}. 
        
        \item \textbf{Smoothness}: The distribution of LOB data representations should be smooth in the input space. The input domain should not be spread in a `salt-and-pepper' manner, which leads to a unclear or complex definition of data validity. Also, the representation should not change dramatically under subtle changes in the market, which results in unstability in the representation as it jumps in the input domains under perturbations. 
    
        \item \textbf{Compatibility}: A good representation should organise information in such a way that can be easily `understood' and processed by machine learning models (instead of humans). This requires the basic assumptions needs to be matched between data representations and learning models. For example, machine learning models usually assumes that each dimension in the input channel contains information from the same source. If not, these models may contain unknown risks due to invalid fundamental settings.
    
\end{itemize}

\section{Towards Robust Representations for Machine Learning Models}

    \subsection{Moving-Window Representation}

        According to the previous example of how adversarial perturbations impact the representation (Sec. \ref{sec:risk}), we learned that vector representations, \textit{i.e.} encoding limit order book volumes in a price-indicated vector, is more robust against market perturbations compared with the compressed representation based on price levels. Instead of using two vectors to represent the bid and ask sides of each limit order book snapshot separately, we choose to use one fixed-length vector centered at the mid price for representing one snapshot. \\
        
        In practice, limit order book data comes in a time-series format. Machine learning models for time-series data normally function on data blocks within a small period of time. This requires the continuity and consistency of representation not only spatially but also temporally. In this paper, we use fixed-size moving windows centred at the mid-price of the current time point to represent a time series of a limit order book. A 2-dimensional window is chosen for the region of interest containing $T$ LOB histories and $2W+1$ continuous price levels stepped by the tick size $\Delta p$. This window provides a vision for limit orders within price range $[p(t) - W\Delta p, p(t) + W\Delta p]$ and within a short history. In the two dimensional matrix $x \in \mathbb{R}^{N \times 2W+1}$, each element $x_{n, i}, n=\{1,...,N\}, i = \{0,...,2W\}$ of the moving window representation indicates the volume of limit orders at price $p(t) - W\Delta p + i$ and at LOB snapshot $t-N+n$. To distinguish ask and bid orders, one can mark $x_{n, i} > 0$ for ask-side limit orders and $x_{n,i} < 0$ for bid side limit orders and the volume size is given by $|x_{n,i}|$. \\
        
        This moving-window representation scheme is easily convertible to other vector or compressed representations. Similar to the other schemes, it sets a \emph{region of interest} to focus on more important information, i.e. limit orders near the current mid-price, rather than distant ones. The choice of window size $W$ reflects the \emph{efficiency} compromise between representational capacity and complexity - large window size include more information about the market but may cause \emph{curse of dimensionality}, while small window size reduces the input space but also limits the view of market to the machine learning model which may potentially decrease the forecasting power. The determination of the region of interest can depend on the tick-size, stock price, accuracy requirement and computational resources of specific tasks. By encoding price information implicitly in the element location and the volume information explicitly in the element value, we now have homogeneity in this moving window representation which should be compatible with the majority of machine learning models including CNNs. Furthermore, the advantage of this disentanglement is obvious - each change to the limit order book would be reflected reasonably in the representation without level shifts and similar LOB snapshots would have similar representations.\\
        
        To identify the smoothness of market representations, one needs to define the `neighbourhood' of a given market state, which remains as an open question. Intuitively, one would consider this `neighbourhood' as those similar market states achievable by making changes that have subtle impacts to the market, e.g. subtle changes to the price or volume of orders. The smoothness of the representation scheme can also be considered spatially or temporally. This moving-window representation implies temporal smoothness - the amplitude of change of representation is proportional to the change in the market. Spatially, subtle changes to the volume would also have minimal impact to the moving window representation. However, subtle changes to the price could lead to very different representations since each volume is associated with a certain price, if the price changes, the value would appear in a different entry.
    
    \subsection{Market Depth Representation}
    
        To improve the spatial smoothness of the moving-window representation, we introduce the market-depth representation, which is an accumulated variation than can be converted directly from the moving-window representation. Each element $x_{n, i}, n=\{1,...,N\}, i = \{0,...,2W\}$ in the accumulated moving window representation is the sum of total volumes up to the corresponding price level on each side in the n-th snapshot. This accumulation of limit orders is referred to as \emph{market depth}, which considers the entire levels of the limit order book. The market depth reflects the market's ability to absorb market orders without having the price affected dramatically by large orders. This market depth representation is a time series aligned to the current mid-price.
            \begin{figure}[!htb]
                \centering
                \includegraphics[width=0.8\textwidth]{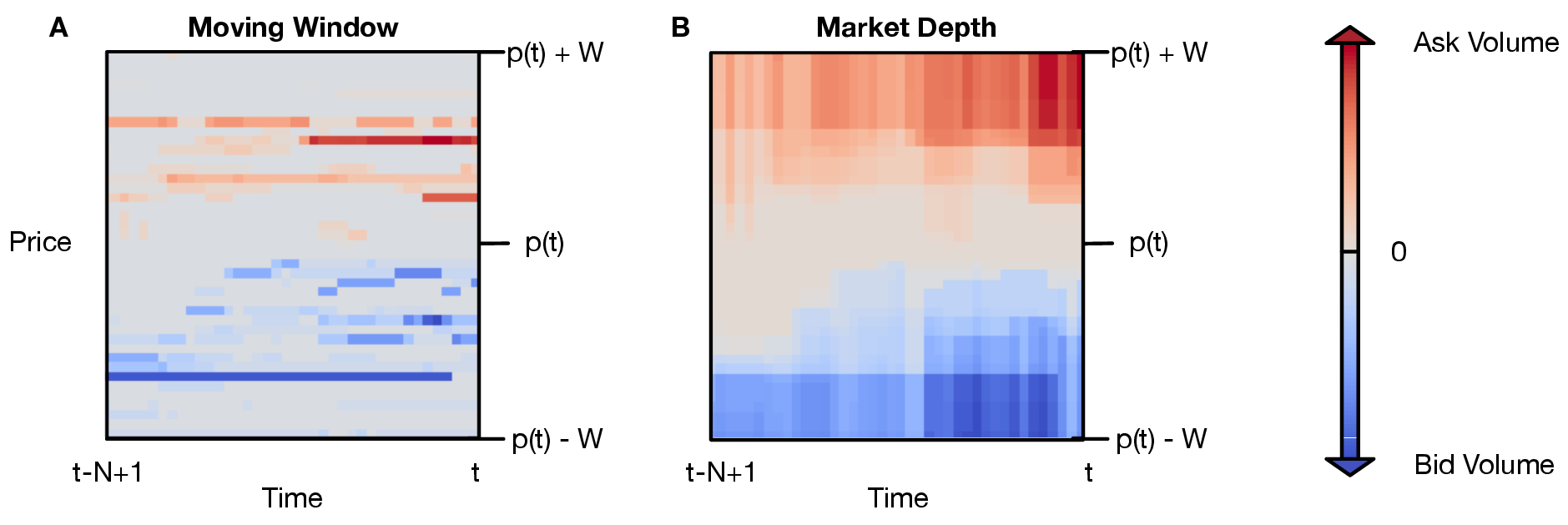}
                \caption{Representation in mid-price-centred Moving Windows. Red/blue represents ask/bid volumes. (\textbf{A}) Moving-window representation. (\textbf{B}) Market-depth representation.}
                \label{fig:new_rep}
            \end{figure}
        
        Figure \ref{fig:new_rep} visualises this moving window representation of a limit order book example $N$ = 40 and $W$= 20. This representation is a reorganisation of limit order book information in a manner to avoid drawbacks of the commonly used compressed representation. First, all numerical variables in this representation are volumes instead of volume-price couples, which avoids the risks of invalidity if disentanglement happens in future black-box models. This representation is also spatially homogeneous since distances between adjacent elements (spatially) are all equal to the tick size. Thus, it does not have the issue of level shifts when empty ticks are filled. Because it includes all the empty ticks within the vision scope, this representation may look sparse, \textit{i.e.} with a considerable amount of zero elements in the representation. Visually, the moving-window representation is more sparse than the market depth representation with more zero entries.

    
    
        \section{Experimental Setup}
    \begin{figure}[!htb]
        \centering
        \includegraphics[width=0.95\textwidth]{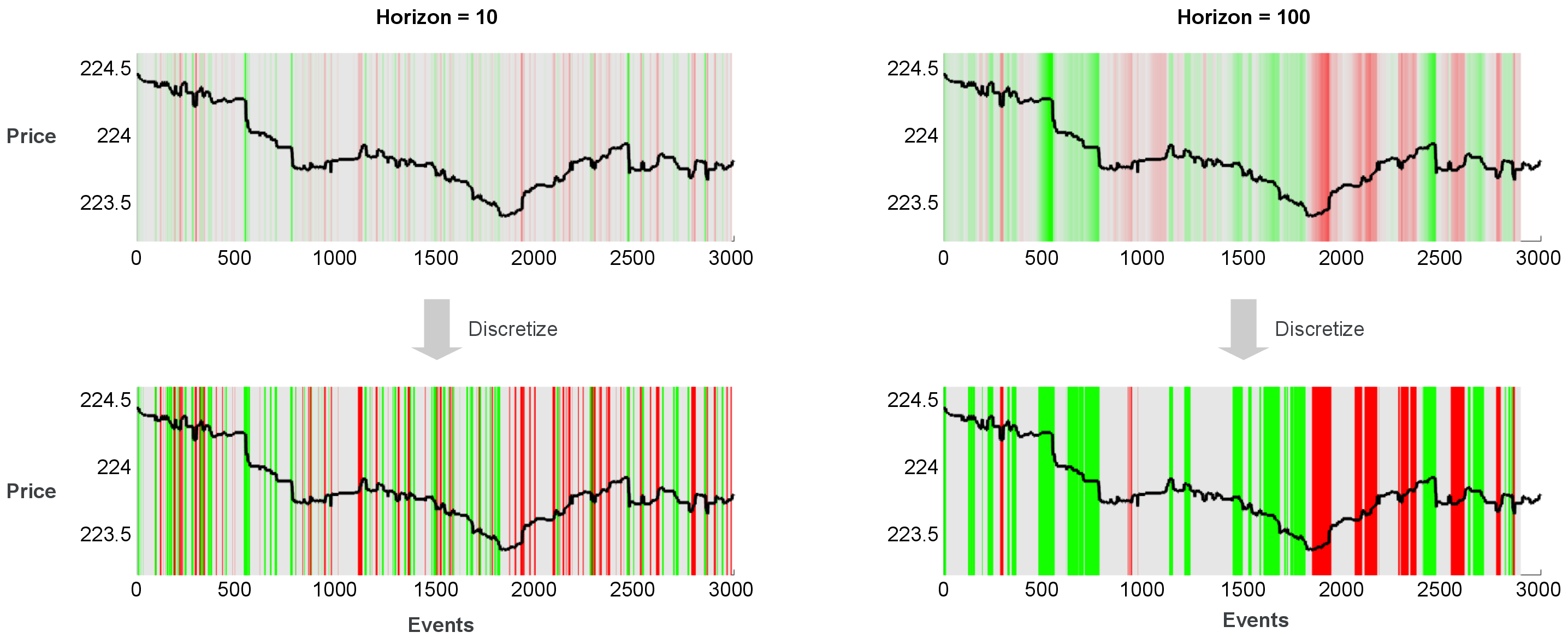}
        \caption{Price forecasting tasks organisation with different horizon. The background colour red/green indicates the up/down trend of the mid-price. \textbf{Top}: the background shading indicates the amplitude of $l_t$ ($l_t > 0$, red) and going down ($l_t < 0$, green).  \textbf{Bottom}: Labels for classification tasks by discretising $l_t$. Grey, red and green represents stationary (0), up (+1) and down (-1).}
        \label{fig:pt}
    \end{figure}
    In our experiments, we use the (weighted) mid-price as our prediction target. Let $p_a^1(t)$ and $p_b^1(t)$ be the best ask and bid prices at time $t$. The \emph{bid-ask spread} is the difference between both prices. The average between the best ask price and the best bid price is referred to as the \emph{mid-price}:
    \begin{equation}
        p(t) = \frac{1}{2} p_a^1(t) + \frac{1}{2}p_b^1(t),
    \end{equation}
    Due to the high stochasticity of market data, the mid-price movement trend is usually defined across a given number of events to reduce noise. This smoothing duration is referred to as the \emph{horizon} of prediction. We follow the definition of \citep{ntakaris2018benchmark} which introduces $l_t$ capturing the movement as the ratio of mid-price change:
    \begin{equation}
        l_t = \frac{m_+(t)-\tilde{p}_t}{\tilde{p}_t},
    \end{equation}
    where $m_+(t) = \frac{1}{k} \sum^k_{i=1}p_{t+i}$ is the smoothed mid-price with prediction horizon $k$.\\
    \\
    Practical price forecasting tasks can be organised as either regression tasks or classification tasks. For regression tasks, the predictive target is the price moving ratio $l_t$ and models need to predict not only which direction the future price is moving towards but also the amplitude of movement (Fig. \ref{fig:pt} \textbf{Top}). For classification tasks, $l_t$ is discretised into labels, which can be achieved in multiple ways. \citep{ntakaris2018benchmark} determines the label based on a threshold $\alpha >0$ and defines `up' ($l_t > \alpha$), `down' ($l_t < - \alpha$) and `stationary' otherwise (Fig. \ref{fig:pt} \textbf{Bottom}). In this way the up and down movement trends are kept symmetric but requires strong prior knowledge for choosing an appropriate $\alpha$ and may cause class imbalance.
    
    \subsection{Dataset}
    
        We use the FI-2010 dataset from \citep{ntakaris2018benchmark} which consists of Level-II LOB data from 5 stocks in the Helsinki Stock Exchange during normal trading hours (no auctions) for 10 trading days. This dataset takes into account 10 price levels on the bid/ask sides of the limit order book, which are updated according to events such as order placement, executions and cancellations. In our experiments, we take into account the history of the LOB snapshots for future price movement prediction. Thus, each input data point is a short time series with input dimension $T \times 40$ where T is the total amount of historical snapshots. The prediction target is the micro-price movement $l_t$ The movement is further categorised into three classes - 1:up ($l_t>0.002$), 2:stationary ($-0.002 \le l_t \le 0.002$), 3:up ($l_t<-0.002$).
    
    \subsection{LOB Representation}
    
        We consider 3 different representations; 1) the compressed representations based on individual price levels that is commonly used in the literature, 2) moving-window representation in a spatial temporal format and 3) market-depth representation as an accumulated version of the moving-window representation. All these representation schemes encode exactly the same information - spatially, price and volume of 10 price levels on each side and temporally, the 10 most recent LOB snapshots as market history. For the compressed representations, the input data dimension is $10\time 40$ and for moving-window and market-depth representations, the input data dimensions are $10 \times 41$. 
    
    \subsection{Price forecasting models}
    
        We apply 5 machine learning models for price forecasting tasks, including 3 basic benchmark models,  \textit{i.e.} linear (logistic regression), multi-layer perceptron (MLP), long short-term memory (LSTM), and 2 deep learning models, DeepLOB \citep{zhang2019deeplob} and temporal convolutional networks (TCNs \citep{bai2018empirical}). The baseline linear model is demonstrated as a multi-class logistic regression. It includes a linear layer with a softmax activation function evaluating the probability of multiple categories. Multilayer perceptrons are generic solutions to machine learning problems without requiring prior knowledge about the spatial-temporal structure of the data. The multi-layer perceptron model we apply has two hidden layers with 100 and 50 neurons respectively with ReLU activation. LSTMs are recurrent neural networks focusing on capturing temporal dynamics of the data. We apply a shallow LSTM model in this paper with only 1 LSTM layer consisting of 20 units. DeepLOB (\citep{zhang2019deeplob}) is a deep learning model for price forecasting combining convolutional neural networks with an LSTM and is designed specifically based on the level-based inputs. The model's hidden layers consist of 3 convolutional layers, 1 inception module and 1 LSTM layer in a sequence, aiming to capture both temporal dynamics and spatial structure of LOB data. Since DeepLOB is not compatible with other LOB representations, we apply temporal convolutional networks to our moving-window and market-depth representations to demonstrate their performance under deep learning settings. TCNs, which capture the spatial-temporal information, are CNN alternatives of conventional RNNs when solving time series problems. The most important structure within TCNs are causal convolutional layers. In this implementation, we stack 3 causal layers (32 channels each) to create a relatively deep but not overly complicated model.

    \subsection{Adversarial Perturbation}
    
        We additionally generate LOB testing datasets with adversarial perturbations by adding orders to limit order books. All 4 input representations now need to represent each perturbed LOB (original LOB + order perturbation) in their own manner. We design 4 perturbation paradigms - the LOB data is not perturbed (`None') and the LOB data is perturbed by placing minimum-size orders to fill the ticks on the ask side only (`ask'), on the bid side only (`bide'), on both the ask and bid side (`both'). The idea of adversarial perturbations is to examine the model's robustness under unexpected subtle perturbations. Robust models should show good generalisation against subtle perturbations and their prediction should not be influenced dramatically even they never seen exact examples like the perturbed ones before. Thus, all these methods are trained with the same FI-2010 training dataset (unperturbed) as how people usually train these models in production but only being tested in perturbation paradigms mimic unexpected adversarial perturbations or even attacks.
    
    \subsection{Performance Evaluation}
    
        Test performance of the machine learning models is measured in the price movement forecasting task described above . Since the test set is unbalanced, we use 2 different metrics to evaluate and compare the performance - Accuracy (\%) and F-score (\%). Accuracy (\%) is measured as the percentage of predictions of the test samples that exactly match the ground truth, which is the unbalanced accuracy score. The F-score is averaged across classes in an unweighted manner to eliminate the influence of data imbalance. All the experiments are repeated 5 times with different random seeds to get an averaged performance with standard deviations. 
        \section{Results}

\subsection{Forecasting Performance}
It can be observed from Table \ref{table:performance} that the ranking of model performance regardless of the input representation or testing paradigm is deep models (DeepLOB, TCN) followed by LSTM followed by MLP and finally Linear. We take the model performance using the compressed representation as an instance. The linear model (accuracy = 52.98\%, F-score = 38.50\%) is not capable of learning complex features either spatially or temporally due to its simplicity. The MLP model with multiple hidden layers in theory can learn arbitrary features. However, the feature extraction in MLP is not effective under limited parameter capacity due to the lack of explicitly defined data structure. Thus, the MLP model performance (accuracy = 60.14\%, F-score = 53.96\%) is much lower than LSTM (accuracy = 60.14\%, F-score = 53.96\%), which has clear definition over the temporal structure during learning. The DeepLOB model builds an additional convolutional architecture on top of a LSTM to enable both spatial and temporal feature extraction and significantly outperforms those relatively simple models (Linear, MLP, LSTM) with accuracy = 77.30\%, F-score = 77.23\%. Similarly, the TCN model also shows leading performance over other comparing methods.\\
            \begin{table*}[!tb]
            \small
                \begin{tabular}{c | c c | c c | c c } 
                     \toprule 
                     \multirow{2}{*}{Perturb.} & 
                     \multicolumn{2}{c}{Compressed (\%)} & \multicolumn{2}{c}{Moving-window (\%)} & \multicolumn{2}{c}{ Market-depth (\%)} \\
                     \cmidrule(lr){2-3} \cmidrule(lr){4-5}  \cmidrule(lr){6-7} 
                     & Accuracy  & F-score  & Accuracy  & F-score  & Accuracy  & F-score  \\
                     \midrule
                     & \multicolumn{6}{c}{Linear} \\
                     \midrule
                     None & 52.98$\pm$0.14 & 38.50$\pm$0.38 & \textbf{59.57$\pm$0.00} & \textbf{53.66$\pm$0.01} & 54.24$\pm$0.00 & 41.12$\pm$0.01 \\
                     Ask & 49.81$\pm$0.09 & 42.47$\pm$0.25 & \textbf{59.57$\pm$0.00} &\textbf{53.66$\pm$0.00} & 54.23$\pm$0.00 & 41.09$\pm$0.00 \\
                     Bid & 51.65$\pm$0.10 &  35.83$\pm$0.16 &\textbf{59.57$\pm$0.00} & \textbf{53.66$\pm$0.01} & 54.24$\pm$0.01 & 41.12$\pm$0.01 \\
                     Both & 49.40$\pm$0.14 & 41.54$\pm$0.04 & \textbf{59.57$\pm$0.00} & \textbf{53.66$\pm$0.00} & 54.24$\pm$0.00 & 41.10$\pm$0.00\\
                     \midrule
                     \midrule
                     & \multicolumn{6}{c}{MLP} \\
                     \midrule
                     None & 60.14$\pm$4.73 &  53.96$\pm$8.41 & 69.52$\pm$0.68 & 67.88$\pm$0.65 & \textbf{71.27$\pm$0.36} & \textbf{69.59$\pm$0.35} \\
                     Ask & 55.75$\pm$2.30 &  47.08$\pm$4.71 & 69.52$\pm$0.68 & 67.88$\pm$0.65 & \textbf{71.27$\pm$0.36} & \textbf{69.59$\pm$0.35} \\
                     Bid & 55.05$\pm$4.08 &  45.74$\pm$8.01 & 69.54$\pm$0.68 & 67.90$\pm$0.65 & \textbf{71.28$\pm$0.36} & \textbf{69.59$\pm$0.35} \\
                     Both & 50.26$\pm$0.89 & 38.88$\pm$5.11 & 69.53$\pm$0.68 & 67.89$\pm$0.65 & \textbf{71.28$\pm$0.36} & \textbf{69.60$\pm$0.35} \\
                     \midrule
                     \midrule
                     & \multicolumn{6}{c}{LSTM} \\
                     \midrule
                     None & 70.74$\pm$0.22 &  68.45$\pm$0.33 & 75.15$\pm$0.40 & 73.89$\pm$0.38 & \textbf{77.46$\pm$0.17} & \textbf{76.18$\pm$0.19} \\ 
                     Ask & 65.85$\pm$1.43 &  63.09$\pm$1.16 & 75.14$\pm$0.40 & 73.89$\pm$0.38 & \textbf{77.46$\pm$0.17} & \textbf{76.18$\pm$0.19} \\
                     Bid & 63.33$\pm$1.58 & 60.39$\pm$1.78 & 75.16$\pm$0.40& 73.88$\pm$0.38& \textbf{77.47$\pm$0.17} & \textbf{76.19$\pm$0.19} \\
                     Both & 57.79$\pm$3.14 & 54.35$\pm$2.74 & 75.15$\pm$0.40 & 73.89$\pm$0.38 & \textbf{77.47$\pm$0.17} & \textbf{76.19$\pm$0.19} \\
                     \midrule
                     \midrule
                     & \multicolumn{6}{c}{DeepLOB \cite{zhang2019deeplob}}
                     \\
                     \midrule
                     None & 77.30$\pm$0.08 & 77.23$\pm$0.10 & / & / & / & / \\
                     Ask & 69.02$\pm$0.33 & 67.42$\pm$0.25 & / & / & / & / \\
                     Bid & 63.94$\pm$0.93 & 60.70$\pm$1.33 & / & / & / & / \\
                     Both & 51.35$\pm$1.74 & 39.59$\pm$3.73 & / & / & / & / \\
                     \midrule
                     \midrule
                     & \multicolumn{6}{c}{TCN}\\
                     \midrule
                     None & / & / & 77.83$\pm$0.33 & 76.47$\pm$0.34 & \textbf{78.81$\pm$0.38} & \textbf{77.29$\pm$0.47} \\
                     Ask & / & / & 77.82$\pm$0.32 & 76.21$\pm$0.34 & \textbf{78.80$\pm$0.38} & \textbf{77.27$\pm$0.47} \\
                     Bid & / & / & 77.83$\pm$0.33 & 76.22$\pm$0.35 & \textbf{78.82$\pm$0.38} & \textbf{77.30$\pm$0.46} \\
                     Both & / & / & 77.83$\pm$0.33 & 76.21$\pm$0.34 & \textbf{78.80$\pm$0.38} & \textbf{77.28$\pm$0.46} \\
                     \bottomrule
                \end{tabular}
                \caption{Price forecasting model performance under data perturbation. Each model is trained with a non-perturbed training set and when testing the model, we apply various data perturbation. None: no perturbation. Ask: perturbation only applied to the ask-side of data. Bid: perturbation only applied to the bid-side of the data. Both: perturbation applied to both ask and bid sides.}
                \label{table:performance}
            \end{table*}
\\
In Table. \ref{table:performance}, we also compare the performance horizontally with different input representations. By just replacing the compressed representation with representations proposed in this paper, the forecasting performance of the same model is boosted by 7\% for Linear (moving-window v.s. compressed), 11\% for MLP (market-depth v.s. compressed) and 7\% for LSTM (market-depth v.s. compressed). Especially, LSTM with the market-depth representation can already reach an approximate level of performance as the much more complex deepLOB model. This means that both the moving-window and market-depth representations are more effective representations than the commonly-used compressed one for machine learning models to learn meaningful features for forecasting future price movements. In general, the market-depth representation demonstrates the best performance among all the representation schemes. This proves from the data-driven perspective that the market depth is important for predicting the evolution of the LOB.
        
\subsection{Robustness under perturbations}
For the compressed representation, we observe a performance decay for all the machine learning models under unexpected perturbations, from the simplest linear model to the more sophisticated DeepLOB model. In addition, the performance decline changes under different types of perturbations. When the perturbation is applied to both sides, the performance decrease becomes more severe - 11\% accuracy decrease on MLP, 12\% on LSTM and over 25\% on DeepLOB. Similar trends can also be viewed for other evaluation metrics. By contrast, our representations both show no obvious performance decay under perturbations in all experiments with different models. \\
\\
From the these performance decay results, we find that DeepLOB, the best model in terms of performance under normal condition as well as the most complicated one, is also the most vulnerable one under perturbation (the largest performance decay). Its predictive accuracy decreases to 47.5\% and the F-score is only 22.2\%, which even underperforms logistic regression. The reason behind this phenomenon may be a combination of various factors. On one hand, the complexity of model is related to overfitting, which may reduce the generalisation ability and become unstable under the perturbation. Also as we mentioned in earlier sections, CNN assumes homogeneous spatial relationship but the compressed LOB representation is obviously heterogeneous, which leads to a mismatch between the data representation and the network characteristics. Once the spatial relationship is further broken due to perturbation, the CNN descriptors may not be able to extract meaningful features and thus cause malfunction of the entire predictor. \\ 
             \begin{figure*}[!tb]
                \centering
                \includegraphics[width=0.9\textwidth]{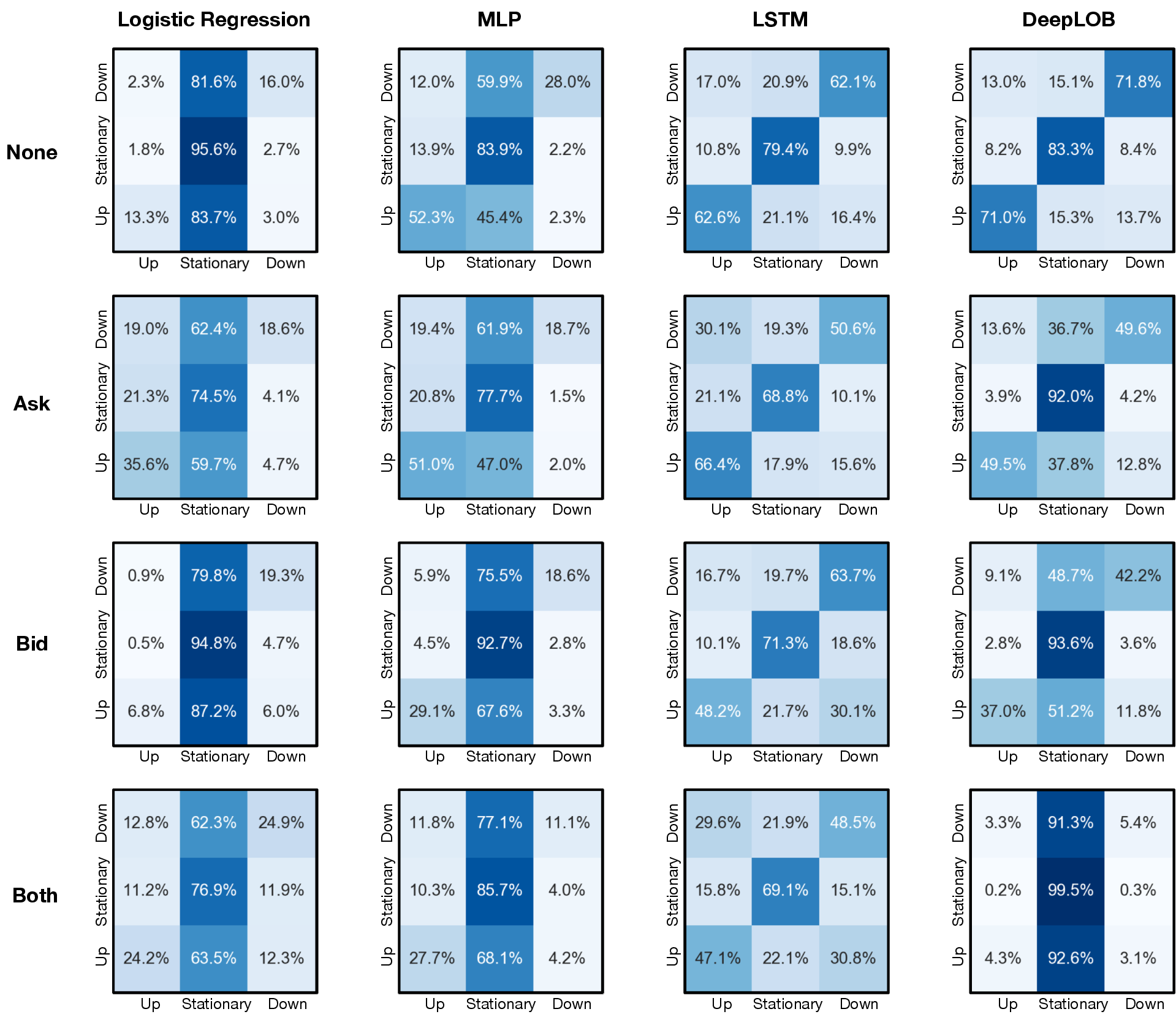}
                \caption{Confusion matrices for corresponding experimental results of the compressed representation in Table. \ref{table:performance}}
                \label{fig:cm}
            \end{figure*}
\\
Figure \ref{fig:cm} provides more details behind the numerical performance metrics in the form of a confusion matrix about the performance decay with the level-based representation. The logistic regression model basically classifies a majority of samples as `Stationary' no matter whether perturbation is applied. Similarly in MLP, about half of the `Up' and `Down' samples are misclassified as `Stationary' ones. Both LSTM and DeepLOB shows confusion matrices with obvious diagonal feature without perturbation - more than half of the samples from each class are classified the same as their true labels. When the perturbation is applied, LSTM shows a performance decrease but still nearly half of samples are correctly classified. DeepLOB, however, fails in the perturbation condition by misclassifying almost all the data to `Stationary' class (see DeepLOB+Both in Fig. \ref{fig:cm}).
        \section{Conclusion \& Future Work}

    In this paper, we discussed the importance of data representations to machine learning models applied to LOB-related tasks and highlighted the drawbacks and risks when using non-robust representations. Further, we proposed new representation schemes that avoid these drawbacks. To reveal this, we implemented price forecasting tasks with multiple benchmark models and data representations. We also designed data perturbation scenarios to not only test the performance but also the robustness of these machine learning models with various representation schemes including the commonly-used level-based representation and our moving window representations. \\ 
    We show that how the information is organised and represented as the input has large impact to the model performance. In our case, by replacing the level-based representation with our moving window representations, performance of the same model increases significantly. This reveals that our moving window representations are more effective and suitable for machine learning models. In addition, the level-based representation brings vulnerability to models even under subtle perturbations, which leads to significant performance decay especially when models are more sophisticated.  Our moving window representations, on the contrary, are almost immune to this perturbation and thus are more stable and reliable. \\ 
    
    Like previous literature, we also show that machine learning models especially deep learning models can be a promising solution to financial problems. Especially, we can adopt existing machine learning solutions (e.g. TCNs, Transformers\cite{lim2021temporal}, Informers\cite{zhou2021informer} etc.) which were designed to solve time-series forecasting problems in various domains. Our future work would also focus on extending robustness to more market-related tasks, especially those utilizing reinforcement learning etc.

    	\spacingset{1}
    
        \section*{Acknowledgements}
            The authors would like to acknowledge our colleagues Vacslav Gluckov, Jeremy Turiel, Rui Silva and Thomas Spooner and  for their input and suggestions at various key stages of the research. This paper was prepared for informational purposes in part by the Artificial Intelligence Research group of JPMorgan Chase \& Co and its affiliates ("J.P. Morgan"), and is not a product of the Research Department of J.P. Morgan. J.P. Morgan makes no representation and warranty whatsoever and disclaims all liability, for the completeness, accuracy or reliability of the information contained herein. This document is not intended as investment research or investment advice, or a recommendation, offer or solicitation for the purchase or sale of any security, financial instrument, financial product or service, or to be used in any way for evaluating the merits of participating in any transaction, and shall not constitute a solicitation under any jurisdiction or to any person, if such solicitation under such jurisdiction or to such person would be unlawful.
        
        \bibliographystyle{chicago}
        \bibliography{references}
	
    \end{document}